\documentclass[aps,prl,twocolumn,letterpaper,superscriptaddress,showpacs]{revtex4}
\usepackage{graphicx}
\usepackage{csquotes}

\begin{document}


\title {\begin{center}
{On the role of chalcogen vapor annealing in inducing bulk superconductivity in Fe$_{1+y}$Te$_{1-x}$Se$_{x}$}
\end{center}}

\author{Wenzhi Lin}
\affiliation{Center for Nanophase Materials Sciences, Oak Ridge National Laboratory, Oak Ridge, TN 37831, USA}

\author{P. Ganesh}
\affiliation{Center for Nanophase Materials Sciences, Oak Ridge National Laboratory, Oak Ridge, TN 37831, USA}

\author{Anthony Gianfrancesco}
\affiliation{UT/ORNL Bredesen Center, University of Tennessee, Knoxville, TN 37996, USA}

\author{Jun Wang}
\affiliation{Center for Nanophase Materials Sciences, Oak Ridge National Laboratory, Oak Ridge, TN 37831, USA}

\author{Tom Berlijn}
\affiliation{Center for Nanophase Materials Sciences, Oak Ridge National Laboratory, Oak Ridge, TN 37831, USA}
\affiliation {Computer Science and Mathematics Division, Oak Ridge National Laboratory, Oak Ridge, Tennessee 37831, USA}

\author{Thomas A. Maier}
\affiliation{Center for Nanophase Materials Sciences, Oak Ridge National Laboratory, Oak Ridge, TN 37831, USA}
\affiliation {Computer Science and Mathematics Division, Oak Ridge National Laboratory, Oak Ridge, Tennessee 37831, USA}

\author{Sergei V. Kalinin}
\affiliation{Center for Nanophase Materials Sciences, Oak Ridge National Laboratory, Oak Ridge, TN 37831, USA}

\author{Brian C. Sales}
\affiliation{Materials Science and Technology Division, Oak Ridge National Laboratory, Oak Ridge, TN 37831, USA}

\author{Minghu Pan}
\thanks{Author to whom correspondence should be addressed. Electronic mail: mhupan@gmail.com .}

\affiliation{School of Physics, Huazhong University of Science and Technology, Wuhan 430074, China}


\begin{abstract}
Recent investigations have shown that Fe$_{1+y}$Te$_{1-x}$Se$_{x}$ can be made superconducting by annealing it in Se and O vapors. The current lore is that these chalcogen vapors induce superconductivity by removing the magnetic excess Fe atoms. To investigate this phenomenon we performed a combination of magnetic susceptibility, specific heat and transport measurements together with scanning tunneling microscopy and spectroscopy and density functional theory calculations on Fe$_{1+y}$Te$_{1-x}$Se$_{x}$ treated with Te vapor. We conclude that the main role of the Te vapor is to quench the magnetic moments of the excess Fe atoms by forming FeTe$_{m}$ (m $\geq$  1) complexes. We show that the remaining FeTe$_{m}$ complexes are still damaging to the superconductivity and therefore that their removal potentially could further improve superconductive properties in these compounds.
\end{abstract}

\maketitle

The interplay between magnetism and superconductivity in cuprate, heavy-fermion and iron-based superconductors has attracted intensive interest of the research community \cite{2000Coleman406,2013Karki110,2012Singh13}. Superconductivity and magnetic ordering are usually considered to be incompatible. In the high-T$_{\textit{c}}$ cuprates, long-range antiferromagnetic order in the parent compounds is suppressed by hole or electron doping before the emergence of superconductivity \cite{Birgeneau2006,2008Zhao7}. 
It has been reported that superconducting states and spin-density-wave (SDW) states are close in the phase diagram \cite{2010Vavilov23}. Recent studies on Ba(Fe$_{1-x}$Co$_{x}$)$_{2}$As$_{2}$ suggest the coexistence of the spin density wave state and superconducting state \cite{2009Laplace80,2009Christianson103}. The nature of the relationship between magnetism and superconductivity is under much debate.
Density functional calculations show strong spin fluctuations in 11-structure iron chalcogenides \cite{2012Singh13}. Despite the competition between magnetism and superconductivity, in the 11-structure system, there is evidence for a coupling between spin excitations and superconductivity \cite{2011Wen74}. Therefore, exploring the role of magnetic impurities offers a pathway to better understand the mechanism of unconventional superconductivity in iron chalcogenides.  

Reported studies have shown the enhancement of superconductivity by annealing iron chalcoginides with excess iron in different types of vapors, such as oxygen and selenium \cite{2013Sun26,2012Hu25,2013Sun82a,2014Zhou83}. Recently, the recovery of bulk superconductivity in annealed Fe$_{1+y}$Te$_{1-x}$Se$_{x}$  sample (under Te vapor) has been reported \cite{2013Koshika82,2013Sun82}. The current lore is that the chalcogen vapor removes excess Fe during annealing. It is timely to investigate this assumption against microscopic measurements such as scanning tunneling microscopy. In fact, spatially resolved studies of impurities are particularly useful for such systems because of the critical role of inhomogeneity in the emergent states. High resolution scanning tunneling microscopy enables direct probing of the interplay between spatial superconducting inhomogeneity and impurities.

Here, we report that by annealing in Te vapor, the magnetic moment of the excess Fe impurities in the iron chalcogenide, Fe$_{1.05}$Te$_{0.75}$Se$_{0.25}$, is reduced.  The moment reduction appears to occur through an electronic hybridization of the excess iron with Te atoms from the vapor that results in the formation of a Fe-Te complex. The formation of this complex results in the recovery of bulk superconductivity. Local structure, electronic properties and the magnetic moment of impurities in the annealed Fe$_{1.05}$Te$_{0.75}$Se$_{0.25}$ sample in Te vapor (hereinafter referred to as \enquote{annealed Fe$_{1.05}$Te$_{0.75}$Se$_{0.25}$ sample}) were investigated in detail by using scanning tunneling microscopy/spectroscopy (STM/S), density  functional theory (DFT) calculations, and bulk measurements. 

Single crystals of Fe$_{1.05}$Te$_{0.75}$Se$_{0.25}$ are grown from the melt using a modified Bridgmann method \cite{2009Sales79,2013Lin7}, and annealed under Te vapor (see the materials and methods section in supplementary S6 for the annealing procedure). X-ray data confirm the good quality of the single crystals and EDX measurements yield an excess iron concentration of about 4-5\%. Furthermore, overall compositions at different spots on crystals before and after annealing were measured using EDX to study the annealing effect. By averaging over 6 spots (regions) on the as-grown and the cleaved shiny surface from the crystal treated in Te vapor (identical analysis geometry), there seems to be a systematic change. The as-grown crystal has an average iron content of 1.057 while for the shiny surface of the crystal treated in Te vapor the average iron content is 1.039. Similarly the as-grown surface has a Te content of 0.718 while the Te-vapor surface has a Te content of 0.743. Apparently, the annealing process does not only remove excess Fe from the sample, but also injects excess Te. 
\begin{figure}[htbp]
\centering
\includegraphics[angle=0,width=3 in]{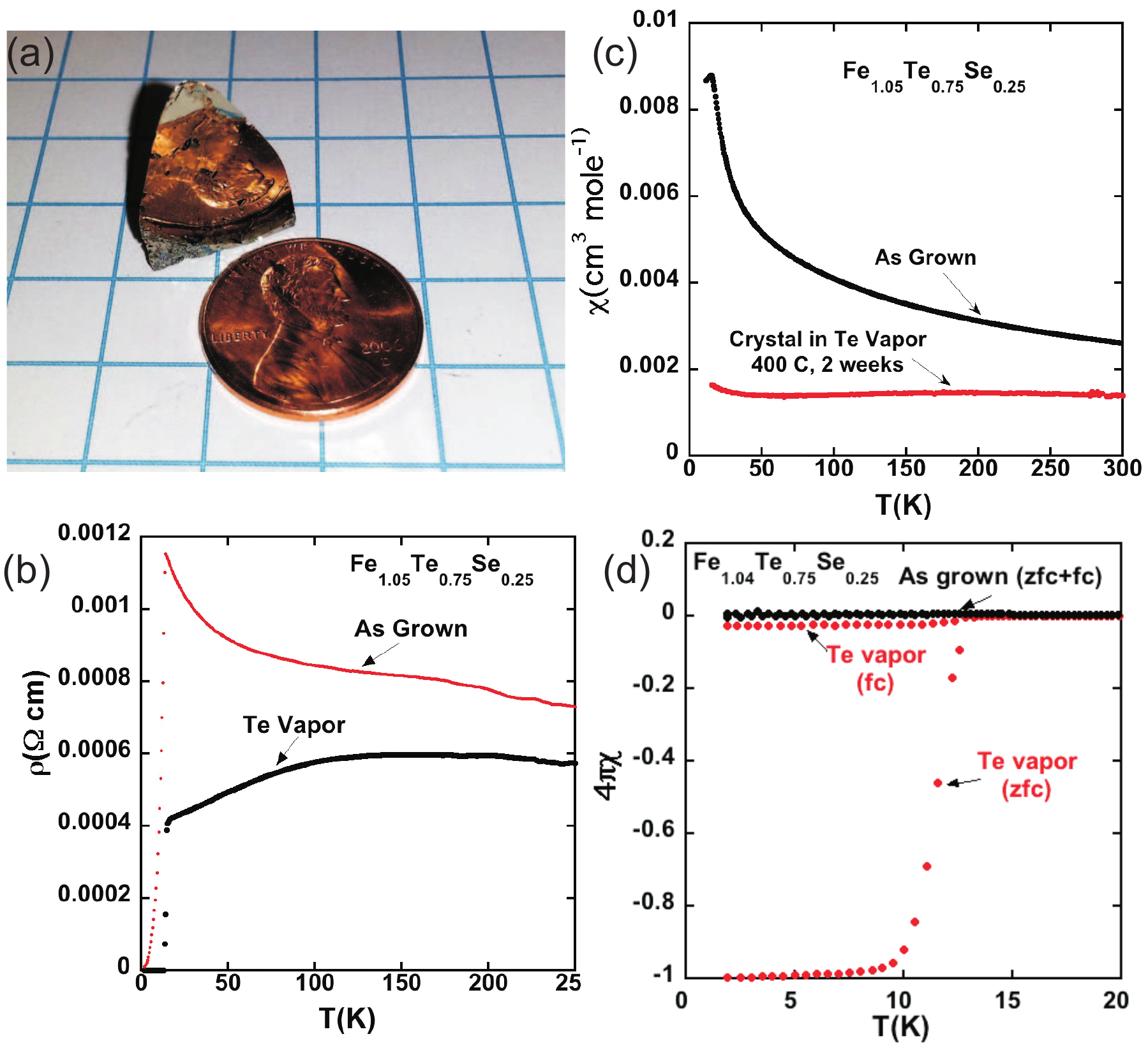}
\caption{(a) Picture of an as-grown Fe$_{1.05}$Fe$_{0.73}$Se$_{0.27}$ sample. (b) Transport measurements of as-grown Fe$_{1.05}$Fe$_{0.75}$Se$_{0.25}$ sample (red curve) and annealed Fe$_{1.05}$Fe$_{0.75}$Se$_{0.25}$ sample under Te vapor (black curve). (c) Magnetic susceptibility measurements of as-grown Fe$_{1.05}$Fe$_{0.75}$Se$_{0.25}$ sample (black curve) and annealed Fe$_{1.05}$Fe$_{0.75}$Se$_{0.25}$ sample under Te vapor (red curve) (d) Zero-field-cooled (zfc) and field-cooled (fc) magnetic susceptibility measurements of as-grown Fe$_{1.04}$Fe$_{0.75}$Se$_{0.25}$ sample (black curves) and annealed Fe$_{1.04}$Fe$_{0.75}$Se$_{0.25}$ sample under Te vapor (bottom red curves) at low temperatures.}
\label{fig:Fig1_BulkMeasurement}
\end{figure}

As an example, Fig. 1a shows an as-grown Fe$_{1.05}$Fe$_{0.73}$Se$_{0.27}$ sample, with the flat and reflective surface perpendicular to the c-axis. Transport and magnetic susceptibility measurements of as-grown crystals of Fe$_{1.05}$Te$_{0.75}$Se$_{0.25}$ or Fe$_{1.04}$Te$_{0.75}$Se$_{0.25}$ show characteristic of filamentary superconductivity (red curve in Fig. 1b, black data in Fig. 1d) \cite{1987Grant58}.  As-grown Fe$_{1.05}$Te$_{0.75}$Se$_{0.25}$ crystals were annealed in Te vapor at 400 $^{\circ}$C for two weeks.  The annealed crystals exhibit a sharp superconducting transition at T$_{\textit{c}}$ $=$ 14 K (Fig. 1b, 1d) in transport, low field magnetic susceptibility data, and heat capacity data (See supplementary Fig. S4). Furthermore, high temperature magnetic susceptibility measurements (Fig. 1c) show that the annealed sample of Fe$_{1.04}$Te$_{0.75}$Se$_{0.25}$ is much less magnetic than the as-grown one. 
\begin{figure}[htbp]
\centering
\includegraphics[angle=0,width=3 in]{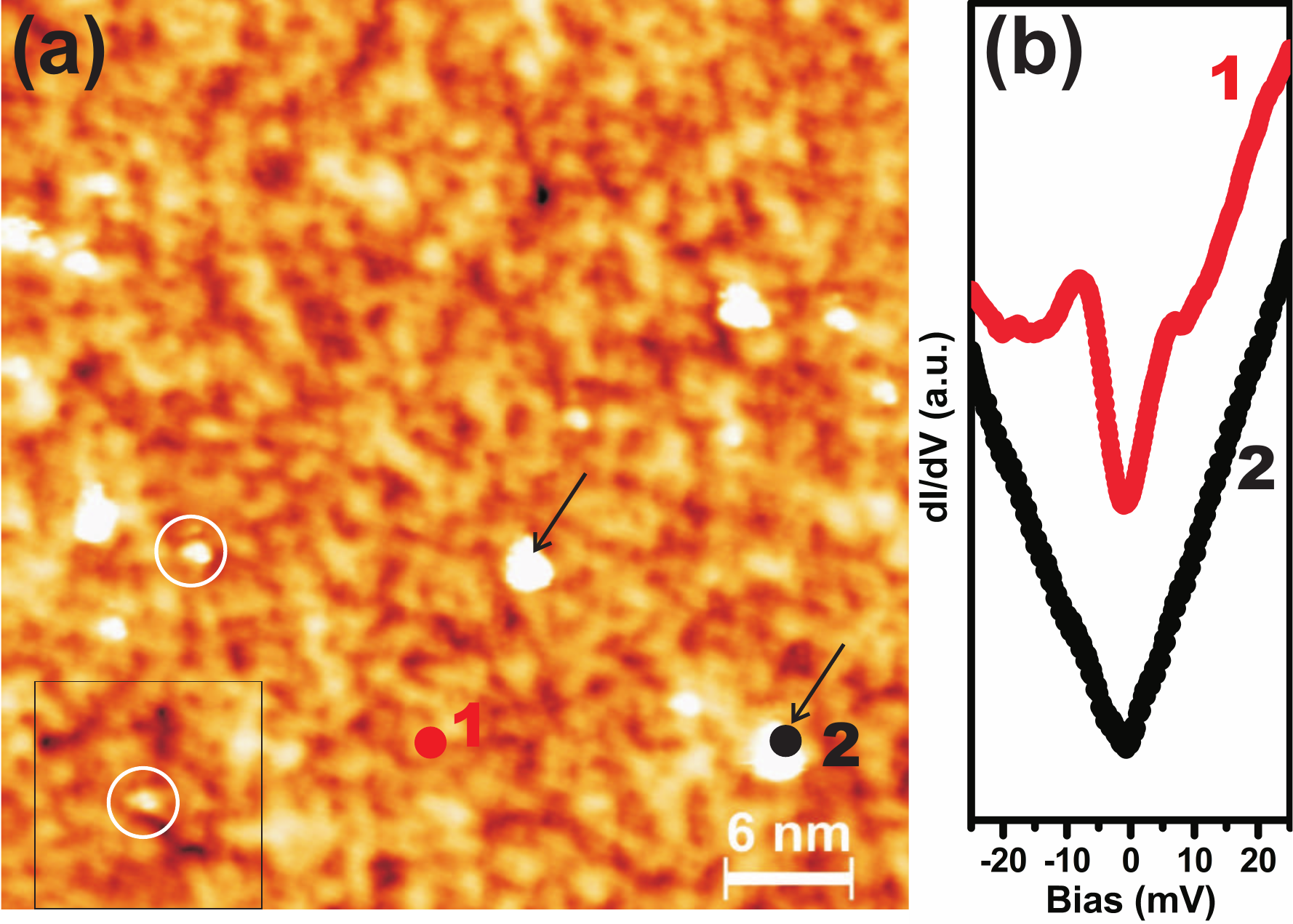}
\caption{(a)  Scanning tunneling microscopy topographic image of the annealed Fe$_{1.05}$Te$_{0.75}$Se$_{0.25}$. The image was taken with a bias of 50 mV and a tunneling current of 100 pA at about 4.3K. There are larger white blobs that are marked with arrows, and relatively smaller bright spots that are marked with open white circles (b) Typical dI/dV spectra taken on the pristine area (red curve) and on the top of an impurity (black curve). The two spectra (1 and 2) correspond to the locations marked by spots 1 and 2 in a. The spectra have been smoothed for better illustration. }
\label{fig:Fig2_LargeArea}
\end{figure}

To study the influence of the impurities on superconductivity, besides bulk superconductivity measurements, we perform local imaging and spectroscopy measurements using STM/S. Annealed Fe$_{1.05}$Te$_{0.75}$Se$_{0.25}$ single crystals were cleaved in ultrahigh vacuum at room temperature (RT) and loaded into the microscope stage to cool down overnight to low temperatures (the microscope stage cooled with liquid helium). The topographic STM images (Fig. 2a, Fig. 3a, and supplementary Fig. S1a) reveal  typical small bright and dark patches, which are attributed to Te/Se clustering on cleaved chalcogenide surfaces \cite{2009Massee80,2011He83}. Furthermore, there are extra bright spots in the images. Previous studies of extra bright spots have been reported on a Fe$_{1+\delta}$Se$_{1-x}$Te$_{x}$ cleaved surface, suggesting that the existence of extra bright spots is likely due to excess iron \cite{2009Massee80}. By examining the impurities and the local electronic properties in combination with theoretical calculations, we are able to clarify the nature of these impurities on the annealed Fe$_{1.05}$Te$_{0.75}$Se$_{0.25}$ sample. 
\begin{figure}[htbp]
\centering
\includegraphics[angle=0,width=3 in]{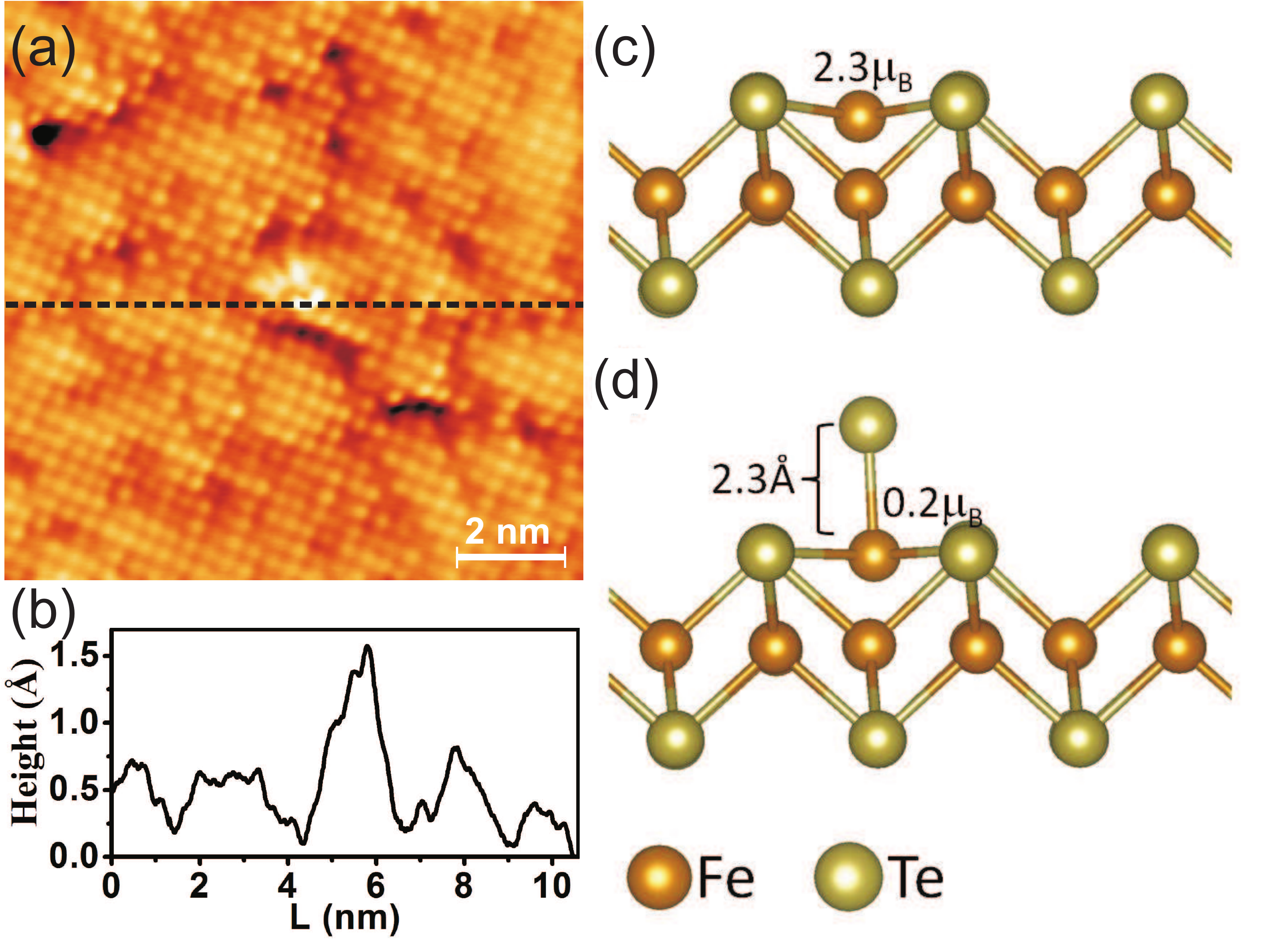}
\caption{(a) Zoomed-in image for the region marked with the square in Fig. 2a reveals an individual impurity (bright spot) as well as surface lattice of the annealed Fe$_{1.05}$Te$_{0.75}$Se$_{0.25}$. The dashed line marks where the height profile in b is extracted. The image was taken with a bias of 50 mV and a tunneling current of 150 pA at about 4.3K. An image flattening process has been applied for better illustration. (b) Height profile along the horizontal dashed line across the bright spot in a, shows the height of impurity is about 1.0$-$1.5 \AA . The height profile was extracted before the image flattening process was applied to a. (c) Relaxed atomic positions obtained from first principles calculations for a single Fe interstitial atom on an FeTe surface.  (d) Relaxed atomic positions obtained from first principles calculations for a single FeTe complex on an FeTe surface. }
\label{fig:Fig3_STMModel}
\end{figure}

We perform scanning tunneling spectroscopy to study the local electronic structure of the impurities. The image in Fig. 2a shows a topographic area of cleaved Fe$_{1.05}$Te$_{0.75}$Se$_{0.25}$ after annealing in the Te vapor. This area displays a relatively high amount of superconductive regions, which is also visualized in superconducting gap map (See supplementary Fig. S3a), and small amount of impurities. The magnetic susceptibility measurement suggests 100\% screening in a zero-field-cooled measurement, which does not imply a 100\% volume fraction. Heat capacity data (See supplementary Fig. S4) indicate that about 65\% of the sample volume is superconducting. The image shows the larger clusters, seen as large white blobs marked with arrows, and relatively smaller bright spots, marked with open white circles. The smaller bright spots still show some structural protrusions, while we observe no structural feature in the larger blobs. It is possible that the compositions in larger blobs are even more complex, and may evolve from the clusters of excess Fe-Te during annealing. Figure 2b shows two typical dI/dV spectra taken on the top of an impurity and the pristine area, respectively. The superconducting coherence peaks vanishes at the impurities (the black curve), while in pristine areas away from the ‘blob’ the superconducting gap and coherence peaks are clearly revealed by inspecting the red curve in Fig. 2b. We also conduct our investigation of local impurities by zooming in the box in Fig. 2a to obtain a high-resolution image of the cluster (Fig. 3a). This topographic image shows a pristine surface with a clear square lattice accompanied by a cluster of impurity in the middle of the image. The line profile (Fig. 4b) along the horizontal dashed line (Fig. 4a) shows a cluster width of about 1.5$-$2 nm with height of about 1.0$-$1.5 \AA\ . 

Density functional theory (DFT) calculations were carried out to investigate these impurities. A single Fe interstitial atom on a FeTe surface (Fig. 3c) prefers a location in the middle of four-chalcogenide atoms on the surface with adsorption energy of 4.12 eV, at a height of  $\sim$0.3 \AA\ below the surface Te-atoms. During annealing in Te-vapor, Te atoms could be introduced into the material, either forming excess Te or binding with Fe interstitial to form FeTe (or FeTe$_{m}$ with m $>$ 1) complex\cite{2013Koshika82,2013Sun82}. The adsorption energy of a single Te-atom on the FeTe surface is 1.20 eV, whereas it is more strongly adsorbed by 1.94 eV on the interstitial Fe-atom on the surface to form a FeTe-complex. This single FeTe complex on the surface of FeTe film has geometry as shown in Fig. 3d.  The FeTe bond length is 2.47 \AA\ . The adsorption energy of such FeTe complex is 2.61 eV, suggesting that it remains strongly adsorbed to the surface, instead of being removed into the gas phase.  The presence of a Te-impurity above the FeTe surface is consistent with the protrusion at the impurity seen in STM experiments.

Most interestingly, the Fe-atoms in FeTe complex are much less magnetic, with a moment of $\sim$0.2 $\mu$B, over 10-times less than $\sim$2.3 $\mu$B seen in an isolated excess Fe-impurity. As seen in supplementary Fig. S3b, the theoretical integrated magnetization density on excess Fe atom is suppressed over the entire energy range when it is bonded to a Te-impurity.  Because of the effect that Te covalently binds with Fe, and reduces its magnetic moment, this provides an explanation for the reduced magnetic scattering for the samples annealed under chalcogenides vapors (or any other atomic gas that can favorably bind with Fe-atom). While the calculations were done on a perfect surface, a covalently bonded FeTe complex would have reduced local moment even when it is adsorbed in the bulk sample. Further, larger FeTe$_{m}$ (m $>$ 1) complexes could also form on the surface, quenching the Fe-moment even further. This would also lead to a reduced magnetic susceptibility, as seen in the experiments. This finding, combining with other evidences, strongly implies that the recovery of superconductivity in annealed sample is most-likely related to the reduced moment of these impurities. During annealing under Te vapor, the excess Fe can strongly attract and react with Te from the vapor. The overall conclusion is that annealing leads to the formation of FeTe$_{m}$ (m $\geq$ 1). The connection between our simulation and our experimental data is on a qualitative level. Qualitatively our simulation shows (a) there is a strong tendency for the excess Fe to attract an excess Te, and (b) the presence of the excess Te will quench the magnetic moment on the excess Fe.
\begin{figure}[htbp]
\centering
\includegraphics[angle=0,width=3 in]{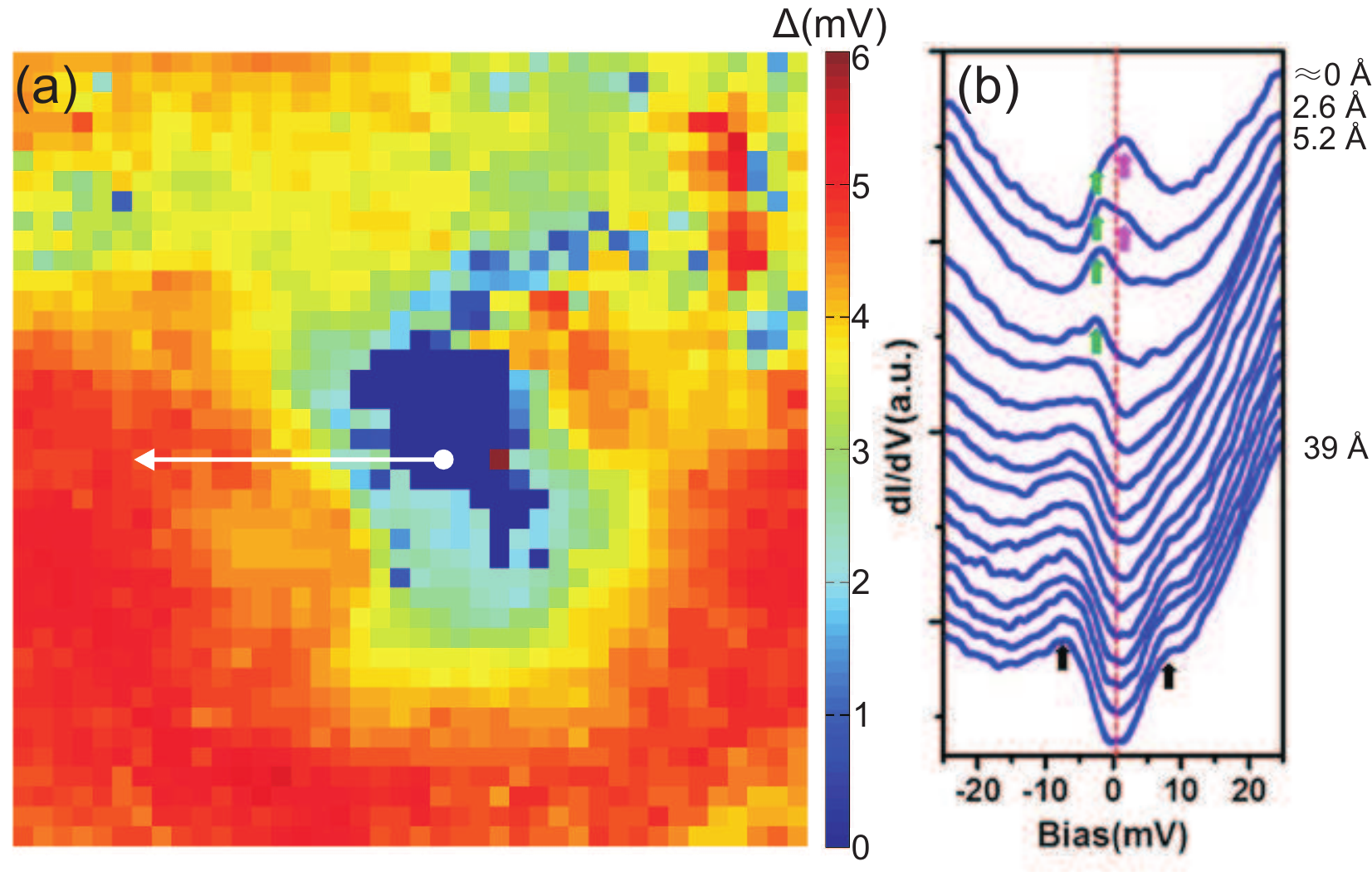}
\caption{ (a) Gap mapping for the same area shown in Fig 3a. This gap mapping is obtained by fitting superconducting gap as described by the references cited in the main text. (Some data points in this map, were corrected manually to 0 value for the reason of the failure of correct fitting by the computer program). (b) The dI/dV spectra measured near and away from the impurity along the arrow mark in Fig 3a, showing the suppression of superconducting gap around the impurity. More interestingly, around the impurity, we also observed that impurity-induced states (indicated by green and pink arrows) appear in the dI/dV spectra with the suppression of the superconducting gap, when tip approaching to the impurity. The top dI/dV curve was taken near the impurity, at the location marked with the white dot in a, which is set as the reference spot (0 \AA\ ). The distance marked for each curve, indicates the linear distance away from the reference spot toward the left side. The set-up conditions for the spectra is V = 50 mV, I$_{\textit{t}}$  = 150 pA, V$_{mod}$ = 0.5 mV. The spectra have been smoothed for better illustration.}
\label{fig:Fig4_STS}
\end{figure}

To analyze the effects of the FeTe$_{m}$  impurities on the superconductivity, we perform scanning tunneling spectroscopy. In Fig. 4a we present a map of the superconducting gap, for the topographic area adjacent to an isolated FeTem impurity shown in Fig. 3a. The gap map was obtained by performing the gap-fitting procedure similar to the method reported in reference \cite{2009Kato80}, as outlined in references \cite{2013Lin7,2012Mitchell86}. The gap map clearly visualizes that superconductivity is suppressed at sites of Fe/Te aggregation. The superconducting gap feature dramatically changes around the impurity. 
Spectra measured around the impurity, reveal either the V-shaped spectra in some areas near the impurity same as the black V-shaped curve in Fig. 2b,  or more interestingly spectra with conductance peaks near the Fermi level as shown in the top curve in Fig. 4b in other areas near the impurity. Both types of spectra indicate the suppression of superconductivity around the impurity. Song \textit{et al}. observed in-gap resonance peak induced inside coherence peak for single iron ad-atom on FeSe \cite{2011Song332}. Here, for Fe$_{1.05}$Te$_{0.75}$Se$_{0.25}$ after annealing in Te vapor, we see the conductance peak around the impurity; however the superconducting coherence peaks disappear. Figure 4a shows that around the impurity, the superconducting gap is greatly suppressed towards 0 meV as indicated in blue. Away from the impurity, the superconducting gap appears with large sizes of about 4$-$5 meV as indicated in red. Previous studies have reported gap values ($\Delta$) for iron chalcogenides, 1.7 meV for Fe(Te,Se) \cite{2010Hanaguri328}, 2.2 meV for FeSe \cite{2011Song332}, and $\sim$ 2$-$4 meV for FeSe$_{1-x}$Te$_{x}$ \cite{2011Fridman72}. The gap sizes of 4$-$5 meV shown in the gap map are near the high value side of the above reported values, which may be due to disorder on our samples. It is clear that the impurity locally suppresses the superconductivity around it. 

As demonstrated in the theory by Abrikosov and Gor$'$kov \cite{1961Abrikosov12}, magnetic impurities suppress superconductivity. In terms of experiments, Yazdani \textit{et al}. reported the experimental STM/STS studies on Mn, Gd, or Ag ad-atoms on an Nb surface, to explore the effects of impurities \cite{1997Yazdani275}. While both magnetic Mn and Gd ad-atoms on the Nb surface enhance conductance around zero bias and locally destroy the superconductivity, nonmagnetic Ag ad-atoms show no effect on the superconducting state on the Nb surface. The latter is a consequence of Anderson’s theorem which states that nonmagnetic impurity does not affect critical temperature in isotropic \textit{s}-wave superconductors \cite{1959Anderson11}. However, in a sign changing superconductor, as the Fe based superconductors are generally believed to be, FeTe$_{m}$ impurity complexes with reduced magnetic moments can still act as strong scatters that locally destroy the supercondcutivity and induce impurity states \cite{2009Zhang103,2009Matsumoto78,2003Zhu67,2011Zhu107}. 

Indeed the presence of in-gap impurity states is revealed in Fig. 4b, which shows the dependence of dI/dV spectra upon lateral distance to the impurity. Far away from the impurity (at about 39 \AA\ ) coherence peaks can be clearly distinguished as marked with the black arrows in the bottom curve of Fig. 4b. Toward the impurity, the coherence peaks are suppressed with a rising structure within the gap which first appears in the electron part of the spectrum at negative bias and then moves to near the zero bias or even the hole side at positive bias (see green and pink arrows at $\sim$ $+/-$ 2 meV). The spectral magnitudes/weights of the in-gap impurity states depend on the distance from the impurity. Such an impurity-induced distance dependent local breakdown of the particle-hole symmetry has been reported before, which exhibits either hole- or electron-like dominant peaks in dI/dV spectra \cite{2003Zhu67,2001Hudson411,2012Li8}. A particularly rapid shift between electron- and hole-like spectral weights is seen by comparing the top two dI/dV spectra in Fig. 4c that are about 2.6 \AA\ away from each other. Such in-gap impurity states with rich spatial and energy properties are not only fascinating, but also contain critical information on the structure of superconducting gap\cite{2006Balatsky78,2009Alloul81,1996Salkola77}. 

The area (shown in supplementary Fig. S2) where the concentration of FeTe$_{m}$ complexes is significantly higher than in the region shown in Fig. 2(a), is found to be non-superconducting. Therefore we can conclude that the effect of the FeTe$_{m}$ complexes is not just limited to suppressing the superconductivity locally, at the nanometer length scale. The FeTe$_{m}$ complexes apparently also conspire together to degrade the superconducting properties in much larger regions of the material. This suggests an intriguing possibility: removing the remaining FeTe$_{m}$ complexes from the superconducting regions may offer a new way to further improve the superconducting properties of the iron chalcogenides. It could very well be that the remaining FeTe$_{m}$ complexes are the limiting factor for the superconducting performance of vapor treated Fe$_{1+y}$Te$_{1-x}$Se$_{x}$ and that their elimination could further improve superconductive properties in these compounds. How the FeTe$_{m}$ complexes could be taken out of the vapor treated iron chalcogenides of course remains to be seen, but the discovery of their existence and the understanding of their destructive influence on the superconductivity have formed the first steps in this direction.

To summarize, we investigate the magnetic and superconducting properties of Fe$_{1.05}$Te$_{0.75}$Se$_{0.25}$ annealed in Te vapor. Using STM/S, theoretical calculations, and bulk measurements, we show that the moments of excess iron atoms are quenched due to the formation of FeTe$_{m}$ (m $\geq$ 1) complexes during the annealing process. Furthermore, the remaining FeTe$_{m}$ complexes still locally suppress the superconductivity. Therefore, we envision that in future experiments there is still room to further improve superconductive properties by not only removing the magnetic moments of the excess Fe impurities, but also the complexes themselves that they form with the Te impurities.

\begin{acknowledgments}

Research was supported (W.L., B.C.S., S.V.K.) by the U.S. Department of Energy, Office of Science, Basic Energy Sciences, Materials Sciences and Engineering Division. This research was conducted (W.L., J.W., M.P., P.G., T.M., T.B.) at the Center for Nanophase Materials Sciences, which is sponsored at Oak Ridge National Laboratory by the Scientific User Facilities Division, Office of Basic Energy Sciences, U.S. Department of Energy. T.B. was supported as a Wigner Fellow at the Oak Ridge National Laboratory. This research used resources of the National Energy Research Scientific Computing Center, a DOE Office of Science User Facility supported by the Office of Science of the U.S. Department of Energy under Contract No. DE-AC02-05CH11231. WSxM software has been used to assist STM data analysis \cite{2007Horcas78}.
\end{acknowledgments}

\end{document}